\documentclass[prd,twocolumn,aps,amsmath,amssymb,nofootinbib,preprintnumbers]
{revtex4}

\usepackage{graphicx}
\usepackage{dcolumn}
\usepackage{bm}
\usepackage{amsmath}
\usepackage{amsfonts}

\def\ls{\mathrel{\lower4pt\vbox{\lineskip=0pt\baselineskip=0pt
           \hbox{$<$}\hbox{$\sim$}}}}
\def\gs{\mathrel{\lower4pt\vbox{\lineskip=0pt\baselineskip=0pt
           \hbox{$>$}\hbox{$\sim$}}}}
\def\drawbox#1#2{\hrule height#2pt

\hbox{\vrule width#2pt height#1pt \kern#1pt
              \vrule width#2pt}
              \hrule height#2pt}

\def\Asym#1#2{\vcenter{\vbox{\drawbox{#1}{#2}
              \kern-#2pt       
              \drawbox{#1}{#2}}}}

\newcommand{\beq}{\begin{equation}}
\newcommand{\eeq}{\end{equation}}

\begin{document}

\title{A stochastic background of gravitational waves from hybrid preheating}

\author{Juan Garc\'\i a-Bellido and Daniel G. Figueroa}

\affiliation{Departamento de F\'\i sica Te\'orica \ C-XI, Universidad
Aut\'onoma de Madrid, Cantoblanco, 28049 Madrid, Spain}

\date{October 14, 2006}

\begin{abstract}
The process of reheating the universe after hybrid inflation is
extremely violent. It proceeds through the nucleation and subsequent
collision of large concentrations of energy density in bubble-like
structures, which generate a significant fraction of energy in the
form of gravitational waves. We study the power spectrum of the
stochastic background of gravitational waves produced at reheating
after hybrid inflation. We find that the amplitude could be
significant for high-scale models, although the typical frequencies
are well beyond what could be reached by planned gravitational wave
observatories like LIGO, LISA or BBO. On the other hand, low-scale
models could still produce a detectable stochastic background at
frequencies accesible to those detectors. The discovery of such a 
background would open a new window into the very early universe.
\end{abstract}
\preprint{IFT-UAM/CSIC-06-46}
\maketitle


According to general relativity, the present universe should be
permeated by a diffuse gravitational wave background (GWB) with a
variety of origins, from unresolved point sources (gravitational
collapse of supernovae, neutron star and black hole coalescense, etc.)
to relic stochastic backgrounds from early universe phase transitions,
inflation, turbulent plasmas, cosmic strings, etc.~\cite{Maggiore}.
These backgrounds have very different spectral shapes and amplitudes
that may, in the future, allow gravitational wave observatories 
like LIGO, LISA, BBO or DECIGO~\cite{Maggiore} to disentangle their
origin. 

There are already a series of constraints on some of these
backgrounds, the most stringent one coming from the large-scale
polarization anisotropies in the Cosmic Microwave Background (CMB),
which may soon be measured by Planck, if the scale of inflation is
sufficiently high~\cite{CMBpol}. There are also constraints coming
from Big Bang nucleosynthesis~\cite{BBN}, as well as from millisecond
pulsar timing~\cite{pulsar}.  Furthermore, it has recently been
proposed a new constraint on primordial GW coming from CMB
anisotropies~\cite{Elena}. Most of these constraints come at very low
frequencies (typically from $10^{-18}$ Hz to $10^{-8}$ Hz), while
present GW detectors work at frequencies of order 1-100 Hz, and
planned observatories will range from $10^{-3}$ Hz of LISA to $10^{3}$
Hz of Advanced-LIGO~\cite{Maggiore}, which could detect GW associated
with early universe phenomena like first-order phase
transitions~\cite{KosowskyTurner,Nicolis}, or cosmic
turbulence~\cite{Turbulence}, if these occur around the electroweak scale.

In this Letter we want to describe a new stochastic background of
gravitational waves that may help open a window into the very early
universe phenomena. Recent observations of the CMB anisotropies seem
to suggest that something like inflation must have occurred in the
early universe. The process by which the energy density driving
inflation is converted into all the radiation and matter we observe
today is called reheating, and corresponds to the true Big Bang of the
Standard Cosmological Model. The first stage of conversion,
preheating~\cite{preheating}, is known to be explosive, and generates
in less than a Hubble time the huge entropy measured today. In chaotic
inflation, the coherent oscillations of the inflaton during preheating
generates, via parametric resonance, a population of highly occupied
modes that behave like waves of matter, which collide among themselves
and whose scattering leads to homogeneization and local thermal
equilibrium. These collisions occur in a highly
relativistic and very asymmetric way, being responsible for the
generation of a stochastic background of gravitational
waves~\cite{TkachevGW,EastherLim} with a typical frequency today of
the order of $10^{7} - 10^{9}$ Hz, corresponding to the present size
of the causal horizon at the end of high-scale inflation. There is at
present no chance to detect such a background, not even by resonant
superconducting microwave cavities~\cite{Picasso}. 

However, there are models like hybrid inflation in which the end of
inflation is sudden~\cite{hybrid} and the conversion into radiation
occurs almost instantaneously. Indeed, since the work of
Ref.~\cite{tachyonic} we know that hybrid models preheat in an even
more violent way than chaotic inflation models, via the spinodal
instability of the symmetry breaking field that triggers the end of
inflation, irrespective of the couplings that this field may have to
the rest of matter. Such a process is known as tachyonic
preheating~\cite{tachyonic,symmbreak} and could be responsible for
copious production of dark matter particles~\cite{ester}, lepto and
baryogenesis~\cite{CEWB}, topological defects~\cite{tachyonic},
primordial magnetic fields~\cite{magnetic}, etc. Moreover, it was
speculated in Ref.~\cite{JuanGW} that in (low-scale) models of hybrid
inflation it might be possible to generate a stochastic GWB in the
frequency range accessible to present detectors, if the scale of
inflation is as low as $H_{\rm inf} \sim 1$ TeV. However, the
amplitude was estimated using the parametric resonance formalism of
chaotic preheating, which may not be applicable in this case.
In Ref.~\cite{symmbreak} it was shown that the process of symmetry
breaking proceeds via the nucleation of dense bubble-like structures
moving at the speed of light, which collide and break up into smaller
structures (see Figs.~7 and~8 of Ref.~\cite{symmbreak}). We
conjectured at that time that such collisions would be a very strong
source of gravitational waves, analogous to the gravity wave
production associated with strongly first order phase
transitions~\cite{KosowskyTurner}. As we will show in this Letter,
this is indeed the case during preheating in hybrid inflation.

Hybrid inflation models~\cite{hybrid} arise in theories of particle
physics with symmetry breaking fields ('Higgses') coupled to flat
directions, and are present in many extensions of the Standard Model,
both in string theory and in supersymmetric theories~\cite{LythRep}.
Inflation occurs along the lifted flat direction, satisfying the
slow-roll conditions thanks to a large vacuum energy $\rho_0$. Inflation
ends when the inflaton $\chi$ falls below a critical value and the
symmetry breaking field $\phi$ acquires a negative mass squared, which
triggers the breaking of the symmetry and ends in the true vacuum,
$\phi=v$, within a Hubble time. These models do not require small
couplings in order to generate the observed CMB anisotropies; e.g. a
working model with GUT scale symmetry breaking, $v=10^{-3}\,M_P$, with
a Higgs self-coupling $\lambda$ and a Higgs-inflaton coupling $g$
given by $g=\sqrt{2\lambda}=0.05$, satisfies all CMB
constraints~\cite{WMAP}, and predicts a tiny tensor contribution to
the CMB polarization.  The main advantage of hybrid models is that,
while most chaotic inflation models can only occur at high scales,
with Planck scale values for the inflaton, and $V_{\rm inf}^{1/4} \sim
10^{16}$ GeV, one can choose the scale of inflation in hybrid models
to range from GUT scales all the way down to TeV scales, while the
Higgs v.e.v. can range from Planck scale, $v=M_P$, to the Electroweak
scale, $v=246$ GeV, see Ref.~\cite{hybrid,CEWB}.

Reheating in hybrid inflation goes through four well defined regimes:
first, the exponential growth of long wave modes of the Higgs field
via spinodal instability, which drives the explosive growth of all
particles coupled to it, from scalars~\cite{tachyonic} to gauge
fields~\cite{CEWB} and fermions~\cite{ester}; second, the nucleation
and collision of high density contrast and highly relativistic
bubble-like structures associated with the peaks of a Gaussian random
field like the Higgs~\cite{symmbreak}; third, the turbulent regime
that ensues after all these 'bubbles' have collided and the energy
density in all fields cascades towards high momentum modes; finally,
thermalization of all modes when local thermal and chemical
equilibrium induces equipartition. The first three stages can be
studied in detailed lattice simulations thanks to the semi-classical
character of the process of preheating~\cite{classical}, while the
last stage is intrinsically quantum and has never been studied in the
lattice.

\begin{figure}[t]
\begin{center}
\includegraphics[width=5.5cm,height=8.5cm,angle=-90]{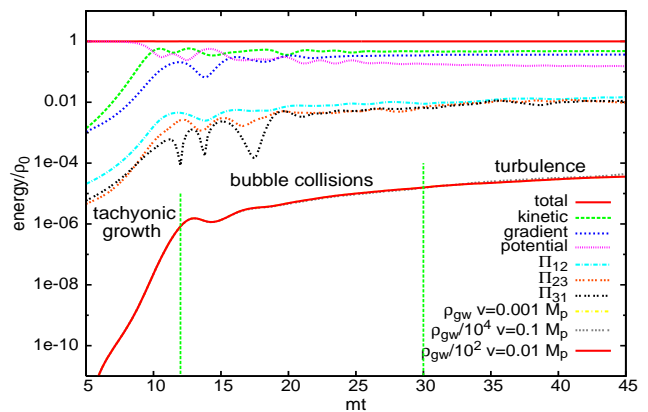}
\end{center}
\vspace*{-5mm}
\caption{The time evolution of the different types of energy (kinetic,
gradient, potential, anisotropic components and gravitational waves
for different lattices), normalized to the initial vacuum energy,
after hybrid inflation, for a model with $v=10^{-3}\,M_P$. One can
clearly distinguish here three stages: tachyonic growth, bubble
collisions and turbulence. }
\label{fig1}
\vspace*{-3mm}
\end{figure}

In this Letter we use lattice simulations to study the generation of
gravitational waves during preheating in hybrid inflation and analyse
the dependence of the shape and amplitude of the spectrum of gravity
waves on the scale of hybrid inflation, and more specifically on the
v.e.v. of the Higgs triggering the end of inflation. Gravitational
waves are represented by a tensor metric perturbation, $g_{\mu\nu}=
\eta_{\mu\nu} + h_{\mu\nu}$, in the transverse traceless (or
radiation) gauge. Its equation of motion in this gauge is $\Box
h_{\mu\nu} = 16\pi G\,T_{\mu\nu}$, with the harmonic gauge condition
$\partial^\mu h_{\mu\nu}=0$ ensured by conservation of the
energy-momentum tensor. In the radiation gauge we can fix $h_{00} =
0$, and the resulting field is the usual tensor gauge-invariant metric
perturbation $h_{ij}$, which satisfies the evolution equation
$h''_{ij} - \nabla^2 h_{ij} = 16\pi G\,\Pi_{ij}$, with $\Pi_{ij}$ the
anisotropic stress tensor, sourced by both the inflaton and Higgs
fields, $\Pi_{ij}=\nabla_i\phi^a\,\nabla_j\phi^a + \nabla_i\chi\,
\nabla_j\chi - 1/3\,\delta_{ij}[(\nabla\phi^a)^2 + (\nabla\chi)^2]$.
We solve the evolution equations of the gravity waves $h_{ij}$
together with those of the other coupled scalar fields in a
discretized lattice, assuming initial quantum fluctuations for all
fields and only a zero mode for the inflaton, following the
prescription adopted in Ref.~\cite{symmbreak}. We also included the GW
backreaction on the scalar fields' evolution via the gradient terms
$h^{ij}\nabla_i\phi\nabla_j\phi$, although for all practical purposes
these are negligible throughout GW production. We then evaluate the
mean field values, as well as the different energy components, see
Fig.~\ref{fig1}. For the energy in gravitational waves we use the
expression $\ (32\pi G)\,t_{\mu\nu} = \left\langle\partial_\mu
h_{ij}^{\rm TT}\, \partial_\nu h^{ij}_{\rm TT}\right\rangle =
{2\over5}\, \left\langle\partial_\mu h_{ij}\,
\partial_\nu h^{ij} \right\rangle$, where the expectation value is
over a region sufficiently large to encompass enough physical
curvature to have a gauge invariant measure of the GW
energy~\cite{Carroll}, and we have expressed the average over the
transverse traceless tensor $h_{ij}^{\rm TT}$ in terms of the average
over $h_{ij}$, the solution of the (traceless) tensor evolution
equation. The fractional energy density in gravitational waves is then
$\rho_{\rm gw}/\rho_0 = 4t_{00}/v^2\,m^2$, which can be used to
compute the corresponding density parameter today (with $\Omega_{\rm
rad}\,h^2\simeq3.5\times10^{-5}$),
$$\Omega_{\rm gw}\,h^2 = \Omega_{\rm rad}\,h^2\,{1\over 8\pi
G\,v^2\,m^2} \left\langle\partial_0 h_{ij}^{\rm TT}\,\partial_0
h^{ij}_{\rm TT}\right\rangle\,,$$ where we have assumed that all the vacuum
energy $\rho_0$ gets converted into radiation, an approximation which
is always valid in generic hybrid inflation models with $v\ll M_P$,
and thus $H\ll m=\sqrt\lambda\,v$. We have shown in Fig.~\ref{fig1}
the evolution in time of the fraction of energy density in GW. The
first (tachyonic) stage is clearly visible, with a slope twice that of
the anisotropic tensor $\Pi_{ij}$. Then there is a small plateau
corresponding to the production of GW from bubble collisions; and
finally there is the linear growth due to turbulence. Note that in the
case that $H\ll m$, the maximal production of GW occurs in less than a
Hubble time, soon after symmetry breaking, while turbulence lasts
several decades in time units of $m^{-1}$. Therefore, we can safely
ignore the dilution due to the Hubble expansion, until the universe
finally reheats and the energy in gravitational waves redshifts like
radiation thereafter.

\begin{figure}[t]
\vspace{-4mm}
\begin{center}
\includegraphics[width=5.5cm,height=8.7cm,angle=-90]{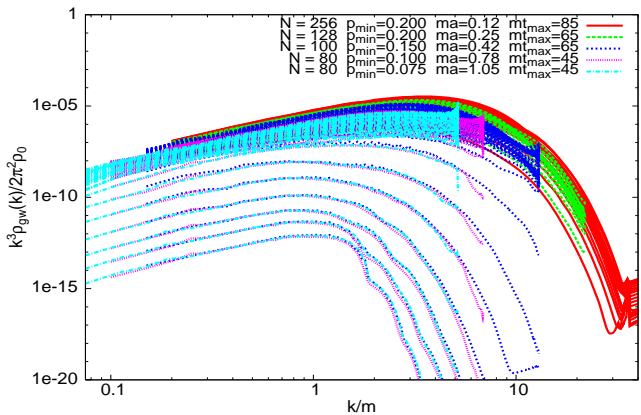}
\end{center}
\vspace*{-5mm}
\caption{We show here the comparison between the power spectrum of
gravitational waves obtained with increasing lattice resolution, to
prove the robustness of our method. The different realizations are
characterized by the number of lattice points (N), the minimum 
lattice momentum (p$_{\rm min}$) and the lattice spacing (ma). The
growth is shown in steps of $m\Delta t = 1$ for the lower
spectra and $m\Delta t = 5$ for the rest.}
\label{fig2}
\vspace*{-3mm}
\end{figure}

We then compute the power spectrum per logarithmic interval in GW by
performing a Fourier transform of the energy density, $\Omega_{\rm
gw}=\int df/f\,\Omega_{\rm gw}(f)$, as a function of the
frequency~$f$, where $\Omega_{\rm gw}(k)=k^3\,\rho_{\rm gw}(k)/2
\pi^2\rho_c$, with $\rho_c$ the critical density today. 
Since gravitational waves below Planck scale remain
decoupled from the plasma immediately after production, we can
evaluate the power spectrum today from that obtained at preheating by
simply converting the wavenumber $k$ into frequency~\cite{TkachevGW},
$$f=6\times10^{10}\,{\rm Hz}\,{k\over\sqrt{H\,M_P}}
=5\times10^{10}\,{\rm Hz}\,{k\over m}\,\lambda^{1/4}\,.$$
We have shown in Fig.~\ref{fig2} the power spectrum of gravitational
waves as a function of wavenumber $k/m$. We have used different
lattices in order to have lattice artifacts under control, specially
at late times and high wavenumbers. We have checked that the power
spectrum of GW follows (turbulent) scaling after $mt\sim40$, and
we can thus estimate the subsequent growth in energy density beyond
our simulations. We have left for a future publication~\cite{alfonso}
the detailed analysis of turbulence in this system.

We will now compare our numerical results with analytical estimates.
The tachyonic growth is dominated by the faster than exponential
growth of the Higgs modes towards the true vacuum~\cite{symmbreak}.
The (traceless) anisotropic strees tensor $\Pi_{ij}$ grows rapidly to
a value of order $k^2|\phi|^2 \sim 10^{-3}\,m^2v^2$, which gives a
tensor perturbation $|h_{ij}^{\rm TT}h^{ij}_{\rm TT}|^{1/2} \sim 16\pi
G v^2 (m\Delta t)^2 10^{-3}$ and an energy density in GW, $\rho_{\rm
gw}/\rho_0 \sim 64\pi G v^2\,(m\Delta t)^2 10^{-6} \sim Gv^2$, for
$m\Delta t \sim 16$.  In the case at hand, with $v=10^{-3}\,M_P$, we
find $\rho_{\rm gw}/\rho_0 \sim 10^{-6}$ at symmetry breaking, which
coincides with the numerical simulations at that time, see
Fig.~\ref{fig1}. The production of gravitational waves in the next
stage proceeds through bubble collisions. Assuming the bubble walls
contain most of the energy density, and since they travel close to the
speed of light~\cite{symmbreak}, it is expected that the asymmetric
collisions will copiously produce GW, like those of a strongly first
order phase transition.  In that case, a quick estimate suggests that
the fraction of energy density is given by~\cite{KosowskyTurner} $\
\rho_{\rm gw}/\rho_0 \sim 1/20 (RH)^2 \sim 8\pi/60\,(Rm)^2 Gv^2 \sim
2Gv^2$, of the same order or slightly larger than the previous stage,
for the typical size of bubbles, $R\sim 3m^{-1}$, upon
collision~\cite{symmbreak}, which again corresponds to what is
observed in the numerical simulations, see Fig.~\ref{fig1}.  The
subsequent turbulent stage~\cite{MichaTkachev,magnetic} is expected to
further produce GW with a spectrum that scales with time in a well
defined manner, see also~\cite{alfonso},
$${k^3\over2\pi^2}{\rho_{gw}(k)\over\rho_0} = 0.2\,Gv^2\,\tau^{1.0}\,
k^2\,\exp(-0.25\,k^2\tau^{-2p})\,,$$
where $\tau=mt$ and $p=1/7$ is the corresponding turbulent 
exponent~\cite{MichaTkachev,magnetic}. This spectrum has a maximum
at $k/m \sim 1$, and falls as $k^2$ for small $k$ until
it reaches the maximum wavelength, $k\sim H$, corresponding to the
minimum frequency today, $f_{\rm min} \sim 5\times10^{10}\,{\rm Hz}\,
\lambda^{1/4}\,v/M_P$. For the case we were considering in our numerical
simulations, with $v=10^{-3}M_P$ and $\lambda\sim g^2\sim0.1$, we find the
power spectrum of Fig.~\ref{fig2}.

\begin{figure}[t]
\begin{center}
\includegraphics[width=5.5cm,height=8.6cm,angle=-90]{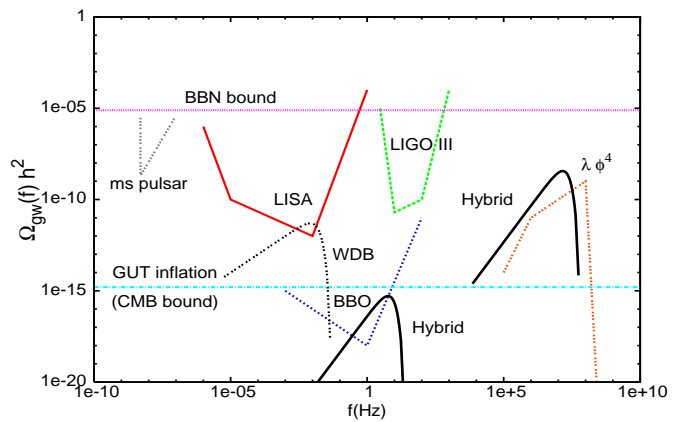}
\end{center}
\vspace*{-5mm}
\caption{The sensitivity of the different gravitational wave
experiments, present and future, compared with the possible stochastic
backgrounds; we include the White Dwarf Binaries (WDB)~\cite{WDB} and
chaotic preheating ($\lambda\phi^4$)~\cite{TkachevGW} for comparison.
Note the two well differentiated backgrounds from high-scale and
low-scale hybrid inflation.}
\label{fig3}
\vspace*{-3mm}
\end{figure}

We have plotted in Fig.~\ref{fig3} the sensitivity of planned GW
interferometers like LIGO, LISA and BBO, together with the present
bounds from CMB anisotropies (GUT inflation), from Big Bang
Nucleosynthesis (BBN) and from milisecond pulsars (ms pulsar).  Also
shown are the expected stochastic backgrounds of chaotic inflation
models like $\lambda\phi^4$~\cite{TkachevGW,EastherLim}, as well as
the predicted background from two different hybrid inflation models, a
high-scale model, with $v=10^{-2}M_P$ and $\lambda\sim g^2\sim0.05$,
and a low-scale model, with $v=10^{-5}M_P$ and $\lambda\sim g^2\sim
10^{-14}$, corresponding to a rate of expansion $H\sim 100$ GeV. The
high-scale hybrid model produces typically as much gravitational waves
form preheating as the chaotic inflation models. The advantage of
low-scale hybrid models of inflation is that the background produced
is within reach of future GW detectors like BBO~\cite{BBO}.

To summarize, we have shown that hybrid models are very efficient
generators of gravity waves at preheating, in three well defined
stages, first via the tachyonic growth of Higgs modes, which act as
sources of gravity waves; then via the collisions of highly
relativistic bubble-like structures with large amounts of energy
density, and finally via the turbulent regime that drives the system
towards thermalization. These waves remain decoupled since the moment
of their production, and thus the predicted amplitude and shape of the
gravitational wave spectrum today can be used as a probe of the
reheating period in the very early universe. The characteristic
spectrum can be used to distinguish between this stochastic background
and others, like those arising from NS-NS and BH-BH coalescence, which
are decreasing with frequency, or those arising from inflation, that
are flat~\cite{SKC}.

For a high-scale model of inflation, we may never see the predicted GW
background coming from preheating, in spite of its large amplitude,
because it appears at very high frequencies, much beyond present
experiments' sensitivities, where no detector has yet shown to be
sensitive. On the other hand, if inflation occured at low scales, even
though we will never have a chance to detect the GW produced during
inflation in the polarization anisotropies of the CMB, we do expect
gravitational waves from preheating to contribute with an important
background in sensitive detectors like BBO. The detection and
characterization of such a GW background, coming from the complicated
and mostly unknown epoch of rehating of the universe, may open a new
window into the very early universe, while providing a new test on
inflation.

{\it Acknowledgments-} We wish to thank Margarita Garc\'\i a P\'erez
and Alfonso Sastre for useful comments on the numerical simulations.
This work is supported in part by CICYT projects FPA2003-03801 and 
FPA2003-04597. D.G.F. also acknowledges support from a FPU-Fellowship
from the Spanish M.E.C.

\end{document}